\documentstyle[aps,aps,epsf]{revtex}

\begin{document}
\newcommand\1{$\spadesuit$}
\tighten
\draft
\twocolumn[\hsize\textwidth\columnwidth\hsize\csname
@twocolumnfalse\endcsname
\title{Quantum Cosmology and Open Universes}
\author{D.H. Coule}
\address{School of Mathematical Sciences, \\
University of Portsmouth, Mercantile House, \\
Hampshire Terrace, Portsmouth PO1 2EG, United Kingdom.\\
e-mail: David.Coule@port.ac.uk}
\author{J\'er\^ome Martin}
\address{DARC, Observatoire de Paris-CNRS UMR 8629, 92195 Meudon Cedex, France. \\
e-mail: martin@edelweiss.obspm.fr}
\maketitle

\begin{abstract}
Quantum creation of Universes with compact spacelike sections that
have curvature $k$ either closed, flat or open, i.e. $k=\pm1,0$
are studied. In the flat and open cases, the superpotential of
the Wheeler De Witt equation is significantly modified, and as a
result the qualitative behaviour of a typical wavefunction differs
from the traditional closed case. In the open case
boundary conditions that include the Tunneling ones are allowed but
not the no boundary choice. Restricting ourselves to
the Tunneling boundary condition, and applying it in turn
 to each of these curvatures, it is shown that quantum cosmology actually suggests that the
Universe be open, $k=-1$. In all cases sufficient inflation $\sim
60$ e-foldings is predicted: this  is an improvement over
classical measures that generally are ambiguous as to whether
inflation is certain to occur.
\end{abstract}
\narrowtext
\vspace{1 cm}]

\section{Introduction}
Quantum cosmology is regarded as a possible way of obtaining
the initial conditions required to start the evolution of a
classical cosmological model, for a general introduction, see Ref. \cite{Kolb}.
 These initial conditions correspond
to a number of arbitrary constants that determine entirely,
in the absence of chaos, the future
evolution of the model. In the simple
Friedman-Lema\^ \i tre-Robertson-Walker (FLRW) metric these constants
determine the amount of matter present, and in the case of a
scalar field source the initial balance of kinetic energy to potential
energy of the scalar field. This determines how the initial
expansion proceeds and whether the strong-energy condition is
 first violated to create an inflationary expansion. One could
 hope to overcome ambiguities in the prediction of whether
 inflation occurs that result from purely classical measures of
 probability \cite{Coule}.

The spatial
curvature $k$ is also another constant to be supplied. Although
with strong-energy satisfying matter e.g. radiation, it will only
dominate at large scale factor to either force the universe to
re-collapse or to keep eternally expanding. But when the strong-energy
condition is violated the curvature instead dominates
at small scale factors. Curvature is therefore especially important
if the universe is assumed to be created in an inflationary state.
But, because  closed $k=1$
models have finite size they have generally been thought to be
most relevant for quantum cosmology as they will have finite
action, see e.g. Ref. \cite{DA}.

The archetypal model that has been considered is the closed
DeSitter space that has at small scale factors a Euclidean or
``forbidden region''. Starting at zero scale factor  one can envision
 tunneling into the classically allowed region. Boundary
 conditions need to be supplied in order to make predictions: the
 two most widely used being the Hartle-Hawking (HH) \cite{HH} and Tunneling
 ones \cite{Tun}. The Tunneling boundary condition can be formulated
 in a number of ways, see \cite{Vil} for a recent discussion,
 but we will generally refer to \cite{Vilenkin1}
  where the Tunneling condition is defined as ``outgoing modes'' only.
   Roughly speaking the HH condition makes use of the Euclidean
 nature of space time to smooth out singularities while the
 Tunneling one corresponds to allowing only outgoing modes,
 making it analogous to quantum $\alpha $ decay of an atom.

To overcome the limitation of only having closed models quantum
creation of bubbles
during any subsequent inflationary phase can enable locally
open regions to form \cite{Cole-DL}.
The use of these so-called Hawking-Turok instantons \cite{HT} are
 essentially making use of the fact that in DeSitter space all
 curvatures are equivalent.
Different slicings of the 5-dimensional DeSitter hyperboloid
correspond to different curvature $k$ when considered as a 4-dimensional
model, see e.g. Ref. \cite{Birr}.  But if the final requirement
is simply an open
or flat universe this would seem, at best, a rather convoluted procedure
for their production.
  Because  open universes can also
be compact \cite{Topo} it seems possible to work directly with
a universe of specific curvature, instead of by necessity
starting with a closed
 universe  that can later by quantum tunneling create locally
  open regions.  This will be the subject of
this paper, for a general
 reference to topology in cosmology see \cite{Topo}. For
  some results this requirement of compactness seems
unnecessary as one can let the volume become arbitrarily large. Or
alternately quantize  the Friedmann equation for the scale factor directly
where  the
 volume factor is a redundant multiplicative factor cf. \cite{HP1}.

Allowing $k=-1$ has a drastic effect on quantum creation scenarios
as the forbidden region that is assumed to be tunneled through
is no longer present. The classical singularity at zero scale
factor is no longer isolated from the larger universe
but instead classical evolution can start
from arbitrarily small size.
\par
 The fact that curvature plays such an
important role in quantum creation scenarios is worrying since the
standard notions of curved space, and their associated metrics, are
 likely to require extensive modifications
  as the quantum gravity epoch is approached. As a first approximation
  to including the effect of quantization of the curvature,
   we will also consider  that the
curvature is initially a quantum (q) variable that is allowed
 to take values within an ensemble. With this assumption
  we will argue that creation is in some sense
``more likely'' if this initial  tunneling is not required which leads one
to conclude that open universes are favoured. A somewhat related
work \cite{mec1} has concluded that
quantum tunneling is favoured if the
 strong-energy condition is only just
being violated. This can be understood since this corresponds to
 making the barrier to be tunneled
through  very shallow and
so easily overcome. We are taking this idea a step further by removing the
barrier entirely.
\par
Because the forbidden region is absent the wavefunctions will be purely
 oscillatory. This is similar to those wavefunctions obtained when the
 Wheeler-DeWitt (WDW) equation corresponding to a classical signature
 change was solved \cite{signa}. It was further found that from notions of
 regularity only a boundary condition analogous to the Tunneling
 boundary condition was allowed.
\par
This article is organized as follows: in section II, we describe the
minisuperspace model for the case where the spacelike sections
have zero, positive or
negative constant curvature and we investigate the general behaviour
of the wavefunctions in such Universes. In section III,
we discuss the choice
of the initial state by making use of regularity requirements
. In section IV, we make physical
predictions using the obtained wavefunctions. In
particular, we compute the probability of creating a Universe with
a particular
spatial curvature and show that it is most likely
 to be an open Universe. The probability for inflation to occur is
 further obtained for any curvature.  We finish
with the conclusions presented in section V.

\section{Description of the model}

We consider the quantization of the following
FLRW metric:
\begin{eqnarray}
\label{defmetric1}
{\rm d}s^2 &=& -N^2(t){\rm d}t^2 \nonumber \\
& &+a^2(t)\biggl(\frac{{\rm d}r^2}{1-kr^2}+
r^2({\rm d}\theta ^2+\sin ^2\theta {\rm d}\varphi ^2)\biggr).
\end{eqnarray}
%In this expression, the scale factor has the dimension of a length.
 It is
well-known that the metric does not fix
the global topology \cite{Topo}. In this article, we will generally
assume that the spacelike sections are
compact (i.e. they have a finite volume) with constant curvature
characterized by $k=0,\pm 1$. Different topologies correspond to
different ranges of variation for the coordinates $(r, \theta, \varphi )$.
The volume of the spacelike hypersurfaces is formally given by:
\begin{equation}
\label{defvol}
v_{k}\equiv \int {\rm d}^3x\sqrt{h},
\end{equation}
where we have written the metric as: ${\rm d}s^2=-N^2(t){\rm d}t^2+
a^2(t)h_{ij}{\rm d}x^i{\rm d}x^j$. Then, the Einstein-Hilbert action
plus the boundary term for
this minisuperspace is given by the following expression:
\begin{equation}
\label{actgrav}
S_{E-H}=\frac{c^3}{16\pi G}v_k\int {\rm d}t Na^3\biggl(\frac{6k}{a^2}
-\frac{6}{N^2}(\frac{\dot{a}}{a})^2\biggr).
\end{equation}
The matter is described by a scalar field whose action
can be written as:
\begin{equation}
\label{actmat}
S_{\phi }=-\frac{v_k}{c}\int {\rm d}tNa^3\biggl(-\frac{1}{2N^2}\dot{\phi }^2
+V(\phi )\biggr).
\end{equation}
In order to pass to the Hamiltonian formalism, conjugate momenta must
be calculated. They are given by:
\begin{equation}
\label{momenta}
\pi _a = -\frac{c^3v_k}{16\pi G}\frac{12a\dot{a}}{N}, \quad
\pi _{\varphi } = \frac{a^3v_k}{c}\frac{\dot{\phi }}{N}.
\end{equation}
>From the last expressions,
 the canonical Hamiltonian can be deduced. It reads:
\begin{eqnarray}
\label{Hamil1}
H_c &=& N\biggl(-\frac{16\pi G}{c^3v_k}\frac{\pi ^2_a}{24a} \nonumber \\
& & +\frac{c}{v_k}\frac{\pi ^2_{\phi }}{2a^3}
-\frac{c^3v_k}{16\pi G}6ka +\frac{v_k}{c}a^3V(\phi )\biggr).
\end{eqnarray}
We are now in a position where the quantization
{\it \`a la Dirac} can be carried out cf. \cite{Kolb,Halli}. It
consists in replacing the two momenta according to the rule:
\begin{equation}
\label{quant}
\pi _a^2\rightarrow -\hbar ^2a^{-p}\frac{\partial }{\partial a}
(a^p\frac{\partial }{\partial a}), \quad
\pi _{\phi }^2 \rightarrow -\hbar ^2\frac{{\partial }^2}{\partial \phi ^2},
\end{equation}
where $p$ takes into account the factor ordering ambiguity. The second rule
is that the action of the operator $\hat{H}_c$ on the wave function
$\Psi (a,\phi )$ gives zero. This leads to the WDW equation:
\begin{eqnarray}
\label{wdw1}
& &\frac{{\rm \partial}^2}{{\rm \partial}a^2}\Psi (a,\phi )+
\frac{p}{a}\frac{{\rm \partial}}{{\rm \partial}a}\Psi (a,\phi )-
\frac{6}{\kappa a^2}\frac{{\rm \partial}^2}{{\rm \partial}\phi ^2}\Psi (a,\phi ) \nonumber \\
& &-\frac{36 v_k^2}{\kappa ^2\hbar ^2c^2}a_0^2(\frac{a}{a_0})^2
[k-(\frac{a}{a_0})^2]\Psi (a,\phi )=0,
\end{eqnarray}
where $\kappa \equiv 8\pi G/c^4$ and
$a_0\equiv [\kappa V(\phi )/3]^{-1/2}$. This equation is not exactly
soluble in its present form so we will make a further assumption.
 Although we still let the
wave function keep a scalar field dependence we will ignore the 2nd
derivative
term w.r.t $\phi$ in the WDW equation. This should be valid
during any  ``slow
roll'' regime where the scalar potential plays the role of an effective
cosmological constant, i.e. $a_0=(3/\Lambda )^{1/2}$. As a
consequence
we find that $a_0^2/(\kappa ^2\hbar ^2c^2)=
3\rho _{\rm Pl}/(512 \pi ^3l_{\rm Pl}^2\rho _{\Lambda })$ where $l_{\rm Pl}$ is
the Planck length and $\rho _{\rm Pl}$ the
Planck energy density, $\rho _{\rm Pl}\equiv c^7/(\hbar G^2)$. $\rho _{\Lambda }$
is the energy density of the effective cosmological
constant, $\rho _{\Lambda }\equiv \Lambda /\kappa $.
 Finally, it is convenient
to work with a dimensionless scale factor expressed in units of
the
Planck length. Therefore we redefine the scale factor according
to: $a \rightarrow l_{\rm Pl}a$. Equation (\ref{wdw1}) can then be rewritten
as:
\begin{eqnarray}
\label{wdw2}
\frac{{\rm d}^2\Psi (a)}{{\rm d}a^2} &+&
\frac{p}{a}\frac{{\rm d}\Psi (a)}{{\rm d}a} \nonumber \\
&-& \frac{27 v_k^2}{128 \pi ^3}\frac{\rho _{\rm Pl}}{\rho _{\Lambda}}
(\frac{a}{a_0})^2[k-(\frac{a}{a_0})^2]\Psi (a)=0.
\end{eqnarray}
The last equation determines the superpotential:
\begin{equation}
\label{superpot}
U(a;k)=\frac{27v_k^2}{128\pi ^3}\frac{\rho _{\rm Pl}}{\rho _{\Lambda }}
(\frac{a}{a_0})^2[k-(\frac{a}{a_0})^2].
\end{equation}
A related  WDW equation has been obtained by using a normalization of the
scale factor to remove any volume factor divergences \cite {HP1}.
In  the following figure, the superpotentials for the three
cases $k=0,\pm 1$ are displayed. Note that
 for $k=0,-1$, the superpotentials
are always negative. This is a crucial difference in comparison with the
 $k=1$ case . This means that when $k=0,-1$ there is no possibility of tunneling
anymore since a zero energy system is always above the superpotential. This
will have  important consequences which are now investigated.
\begin{figure}
\begin{center}
\leavevmode
\hbox{%
\epsfxsize=8cm
\epsffile{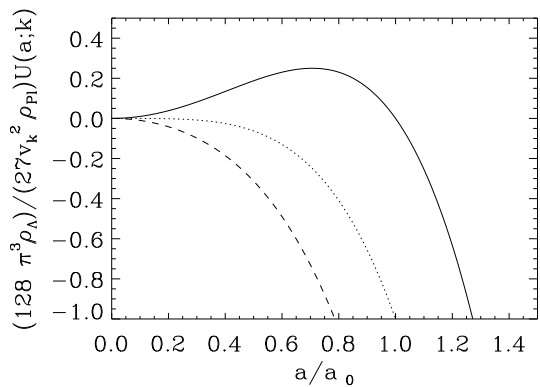}}
\end{center}
\caption{Superpotentials for different values of $k$.
The full line represents
the case $k=1$, the dotted line the case $k=0$ and the dashed
line the case $k=-1$.}
\label{superpotential}
\end{figure}
In order to obtain exact solutions of the WDW equation
we choose to work with the factor
ordering given by $p=-1$. This permits to work with
analytical exact solutions. Introducing the variable $z(a;k)$ defined by
cf.\cite{Vilenkin1}:
\begin{equation}
\label{defz}
z(a;k)\equiv (\frac{3}{8\pi })^{4/3}v_k^{2/3}(\frac{\rho _{\rm Pl}}{\rho _{\Lambda }})
^{2/3}[k-(\frac{a}{a_0})^2],
\end{equation}
the general solution can be expressed in terms of Airy functions
of first and second kind \cite {Abro}:
\begin{equation}
\label{solpsi}
\Psi (a;k)=\frac{\alpha Ai[z(a;k)]+\beta Bi[z(a;k)]}
{\alpha Ai[z(0;k)]+\beta Bi[z(0;k)]}\equiv \frac{N(a;k)}{D(k)},
\end{equation}
where the coefficients $\alpha $ and $\beta $ are arbitrary complex numbers
determined by the choice of a state for the wave function of the
Universe.
\par
We can make some comments at this point. The presence of the
denominator in the previous equation comes from the requirement
that the wave function be regular everywhere in the minisuperspace. Recall
that the wave function depends on the scalar field and the approximation
made previously was only adopted for computational convenience. Let us
therefore re-consider the full WDW equation given
by formula (\ref{wdw1}). It is convenient to define the quantity
 $\alpha \equiv \ln (a/a_0)$. Then the WDW equation
becomes:
\begin{eqnarray}
& & \frac{{\rm \partial }^2\Psi (\alpha ,\phi )}{{\rm \partial }\alpha ^2}+
(p-1)\frac{{\rm \partial }\Psi (\alpha ,\phi )}{{\rm \partial }\alpha }
-\frac{6}{\kappa }\frac{{\rm \partial }^2\Psi (\alpha ,\phi )}{{\rm \partial }\phi^2}
\nonumber \\
& &-\frac{27v_k^2a_0^2}{128\pi ^3}\frac{\rho _{\rm Pl}}{\rho _{\Lambda }}
e^{4\alpha }[k-e^{2\alpha }]\Psi (\alpha ,\phi )=0.
\end{eqnarray}
This equation is separable. If we define $\Psi (\alpha ,\phi )\equiv
e^{\frac{1-p}{2}\alpha }f(\alpha )g(\phi )$ and if $\lambda $ is the
separation constant then the solution can be written as:
\begin{equation}
\Psi (\alpha ,\phi )=\int c(\lambda )e^{\frac{1-p}{2}\alpha }
f_{\lambda }(\alpha )
e^{i\sqrt{\frac{\kappa }{6}}\lambda \phi }{\rm d}\lambda ,
\end{equation}
where $c(\lambda )$ are {\em a priori} arbitrary coefficients and the
function $f_{\lambda }(\alpha )$ satisfies the equation:
\begin{eqnarray}
\label{eqf}
\frac{{\rm d}^2f_{\lambda }(\alpha )}{{\rm d}\alpha ^2} &+& \biggl(\lambda ^2
-\frac{(1-p)^2}{4} \nonumber \\
&-&\frac{27v_k^2a_0^2}{128\pi ^3}\frac{\rho _{\rm Pl}}{\rho _{\Lambda }}
e^{4\alpha }[k-e^{2\alpha }]\biggr)f_{\lambda }(\alpha )=0.
\end{eqnarray}
Note that when considering quantum wormhole solutions with massless
scalar fields the oscillatory divergence can be regulated by a proper
choice of the coefficients $c(\lambda )$ \cite{HP2,Garay,stiff}. But we will rather
consider single wavefunction like solutions for which such a scheme
is absent \cite{Mijic}. When the scale factor
becomes small, $a\ll a_0$, Eq. (\ref{eqf}) can be
readily
solved. The solution is $f_{\lambda }(\alpha )\approx e^{i\omega _{\lambda ,p}\alpha }$
where $\omega _{\lambda ,p}$ can be expressed as:
\begin{equation}
\omega _{\lambda ,p}=\sqrt{\lambda ^2-\frac{(1-p)^2}{4}}.
\end{equation}
In particular when $p=-1$, i.e. the factor ordering adopted in this
article, one has $\omega _{\lambda ,p=-1}=\sqrt{\lambda ^2-1}$. We are now
in a position where the regularity of the wavefunction in the vicinity
of $a=0$ can be studied. It crucially depends on the value of the
separation constant, see also Ref. \cite{Mijic} where a similar
treatment has been performed. If $\lambda \in ]-1,1[$ then the wavefunction
is regular as the scale factor goes to zero whereas otherwise the
wavefunction exhibits rapid oscillations. For these
values of $\lambda $ the wavefunction  remains finite.
Nevertheless, the ``wiggliness'' of the wavefunction is also a kind
of singularity and it seems reasonable not to consider this
possibility . In this paper, we
restrict ourselves to the case $\lambda =0$ for which the
wavefunction is regular at $a=0$. This means that $\Psi (a=0,\phi )$
does not depend on $\phi $ and it justifies the presence of the denominator $D(k)$
in Eq. (\ref{solpsi}), see also Refs. \cite{Vilenkin1}. It is clear from the previous discussion
that this is not the most general case since there exists non vanishing values
of $\lambda $ such that the wavefunction is regular at the origin. However, to
our knowledge, there does not exist
a general study of the boundary conditions for an arbitrary value of the
separation constant. Moreover, it turns out that the two most widely discussed
choices of boundary conditions, i.e. Hartle-Hawking and Vilenkin states, picked
out the value $\lambda =0$ as  will be demonstrated below, see also
Ref. \cite{Mijic} (in these two cases, $D(k)$ is responsible for the
appearance of the factors $e^{\pm 1/V(\phi )}$ in the
probability density functions). So
as a  first approach it seems reasonable to
consider the case $\lambda =0$ only. Let
us emphasize that Eq. (\ref{solpsi}) allows
us to study all the boundary conditions such that  $\Psi (a=0,\phi )$
is independent of the scalar field and not only the Hartle-Hawking and Vilenkin
states.
\par
We should distinguish between the sort of singularity discussed above, present for
example with a massless scalar field case \cite{HP2}, which would
tend to be displayed by a rapid
oscillation in the wavefunction as $a\rightarrow 0$ (recall kinetic
energy $\sim$ ``wiggliness'' of wavefunction, see eg.\cite{Rob}) if not
regulated  and those considered by  Ref.~\cite{australia} which
are related to the choice of the factor ordering. There, for the closed
case, it was claimed that only the no-boundary
state is regular in the limit $a\rightarrow 0$ for any choice
of $p$. It was also shown that the Tunneling solution is regular
in this limit only if $p<1$. This behaviour was caused by
the wavefunction having growing and decaying, actually
Modified Bessel functions, solutions in the forbidden
region. Typically the required Tunneling solution diverges like $\Psi
\sim a^{(1-p)/2}$ in the limit $a\rightarrow 0$. However, this sort of
divergence is present universally. Solving Eq. (\ref{wdw2}) for
$k=\Lambda=0$ gives the solution
\begin{equation}
\label{solflat}
\Psi(a) = c_1+c_2 a^{1-p},
\end{equation}
with $c_1$ and $c_2$ arbitrary constants. So the factor ordering divergence
remains even in flat empty space. These divergences appear intrinsic to these models
and one can
conceive that they should be removed by a renormalization scheme.
It was further suggested, in the context of quantum wormhole that
such factor ordering divergences are of no great
concern \cite{Kim}. In practice, as $a \rightarrow 0$, one should anyway
include a more realistic strong-energy satisfying matter
component. As emphasized by Gott and Li \cite{GL} such a matter
source should be expected if only of a size due to quantum `` zero
point'' fluctuations. One can think of this as being because the
scalar potential has a ``fuzziness'' due to quantum uncertainty in the
limit $a\rightarrow 0$. For the case of radiation, given here by a
parameter A the WDW equation is then given by, see e.g. \cite{Rad}
\begin{equation}
\label{wdwradiation}
\frac{{\rm d}^2\Psi(a)}{{\rm d}a^2}
+\frac{p}{a}\frac{{\rm d}\Psi(a)}{{\rm d}a}+A\Psi(a)=0
\end{equation}
and now with solution,
\begin{equation}
\label{solwdwrad}
\Psi (a)= c_1a^{\frac{1-p}{2}}J_{\frac{p-1}{2}}(\sqrt{A}a)
+c_2a^{\frac{1-p}{2}}Y_{\frac{p-1}{2}}(\sqrt{A}a),
\end{equation}
where $J_{(p-1)/2}$ and $Y_{(p-1)/2}$ are Bessel functions of order
$(p-1)/2$ \cite{Abro}. But
as pointed out by \cite{Wilt99} the Y Bessel function
still diverges for $p\geq 1$ and this term is part of the Tunneling
boundary condition's solution. However, the presence now of oscillatory
behaviour complicates the adoption of the HH boundary condition.
It sometimes is given by the prescription ``outgoing'' plus
``ingoing'', if so it would also suffer this same divergence, although if
it was simply the J Bessel function term it would be normalizable
\cite{Wilt99}. But regardless
of such considerations for the chosen $p=-1$ case the no-boundary state
as well as the Tunneling state are immune from factor
ordering divergences and are {\em a priori} both allowed.
\par
Let us now turn to the study of the behaviour of the wave functions
given in Eq. (\ref{solpsi}). The case $k=1$ is standard. We choose
the spacelike sections to be spheres although of course other
possibilities are allowed. This means that $v_1=2\pi ^2$. The evolution
of the Universe can be viewed as the motion of a fictitious particle with
zero energy in the potential given by Eq. (\ref{superpot}). The particle starts
to the left of the potential and can proceed to tunnel through the barrier. While
in the region $a<a_0$ the wavefunction exponentially decays.
Then, when the scale factor is beyond the barrier $a>a_0$, the wavefunction
becomes oscillatory. This behaviour is illustrated in the following
figure\footnote{In this article, the figures have been obtained using
the Mathematica (version 3.0) and IDL software packages} for
the boundary conditions $\alpha =1$ and $\beta =i$, which correspond to the
Tunneling wavefunction \cite{Vilenkin1} .
The value of $\rho _{\Lambda }$ is chosen such
that $\rho _{\rm Pl}/\rho _{\rm \Lambda}
=1000$. Since we have $a_0^2=(3/8\pi)\rho _{\rm Pl}/\rho _{\Lambda }$, this
corresponds to a dimensional $a_0$ equal to $\approx 10.9l_{\rm Pl}$. This
value can be thought of as being the size at  which the Universe is first
created.
\begin{figure}
\begin{center}
\leavevmode
\hbox{%
\epsfxsize=8cm
\epsffile{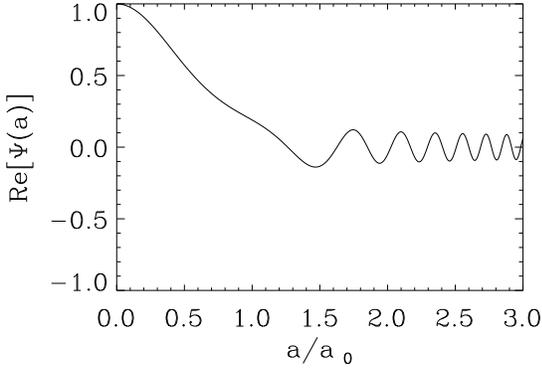}}
\end{center}
\caption{Real part of the Vilenkin wavefunction for $k=1$ and $v_1=2\pi ^2$. The
value $\rho _{\rm Pl}/\rho _{\Lambda }=10$ has been chosen rather than the more
realistic value $\rho _{\rm Pl}/\rho _{\Lambda }=1000$ only for the sake
of illustration. }
\label{wavefunctionk=1}
\end{figure}

The cases $k=0,-1$ are very different. The fictitious particle
is now always above the
potential which is now negative. Tunneling is no longer
 required as classical evolution is possible. As
a consequence the wave function always exhibits oscillatory behaviour. To
go further, we need to know the topology of the spacelike sections. The
smallest three-hyperbolic manifold is not known. The two smallest spaces known
are the Weeks space \cite{Weeks} and the Thurston space \cite{Thurs}. The volume of the
first one is $\approx 0.94 $ and of the second one is $\approx 0.98 $. The results
presented here do not depend crucially on the volume of the spacelike sections
provided they are of the same order of magnitude. Therefore we choose to work
with the Weeks space and consequently we take $v_{-1}\approx 0.94$. In the flat
case the volume $v_0$ is arbitrary. For definiteness, we choose $v_0=1$. The
two following figures show the wavefunction for the cases $k=0$ and $k=-1$
respectively. The boundary conditions chosen
 are the same as previously. It can
be seen that the wavefunctions indeed oscillate even before the scale factor reaches
the value $a_0$.
\begin{figure}
\begin{center}
\leavevmode
\hbox{%
\epsfxsize=8cm
\epsffile{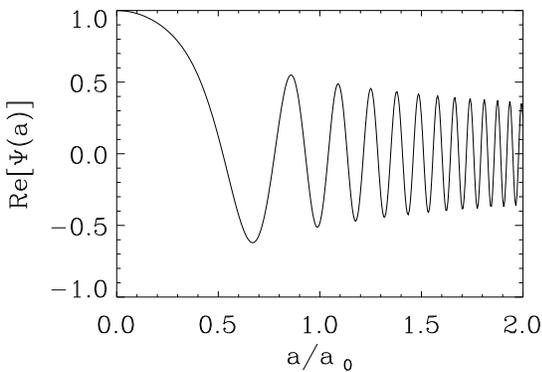}}
\end{center}
\caption{Real part of the Vilenkin wavefunction for $k=0$, $v_0=1$ and
$\rho _{\rm Pl}/\rho _{\Lambda }=1000$.}
\label{wavefunctionk=0}
\end{figure}
It can also be noticed that the oscillations in the region $a<a_0$ are
more pronounced for the  $k=-1$ case  than for when $k=0$. This is due
to the different shapes of the superpotential. The particle
can roll faster down the steeper $k=-1$ superpotential compared to
the $k=0$ case.
\begin{figure}
\begin{center}
\leavevmode
\hbox{%
\epsfxsize=8cm
\epsffile{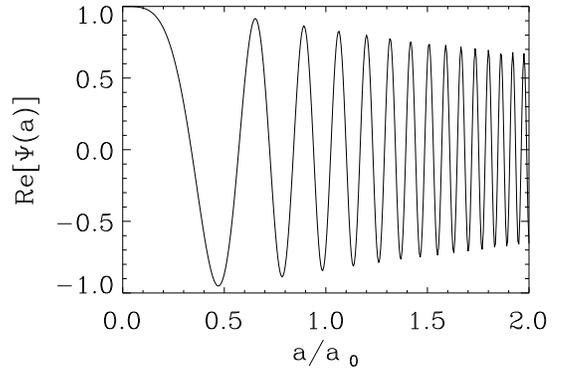}}
\end{center}
\caption{Real part of the Vilenkin wavefunction for $k=-1$,
$v_{-1}\approx 0.94$ and $\rho _{\rm Pl}/\rho _{\Lambda }=1000$.}
\label{wavefunctionk=-1}
\end{figure}
The fact that the wavefunctions behave very differently for different values
of $k$ also has an impact on the choice of the quantum state. Studying this
question is the purpose of the next section.
\section{Measure and the initial state}

To be able to make predictions and to calculate probabilities, we need a
suitable
measure. If one chooses a surface in the minisuperspace perpendicular
to the ``$a$'' direction, then the component of the current associated
with the WDW equation through this surface is given by \cite{Halli}:
\begin{equation}
\label{defcurrent}
j=\frac{i}{2}a^p
(\Psi ^*{\rm \partial}_{a}\Psi-\Psi {\rm \partial }_{a}\Psi ^*),
\end{equation}
where, in our case, $p=-1$. As is well-known this current is not positive
definite since the signature
of the (mini)-superspace is Lorentzian. However, in the WKB
approximation,
where
the wave function can be written as $\Psi \sim Ce^{iS}$, this current
becomes positive definite and permits the calculation of conditional
probabilities \cite{Vilenkin2}.
\par
Let us first calculate the behaviour of the numerator of Eq. (\ref{solpsi})
when the scale factor becomes large. We have $\lim _{a\rightarrow +\infty }
z(a;k)=-\infty $ for any value of $k$. Using the asymptotic behaviour of the
Airy functions, we find:
\begin{eqnarray}
\label{wkbnum}
\lim _{a\rightarrow +\infty }N(a;k) &=&
\frac{[-z(a;k)]^{-\frac{1}{4}}}{2\sqrt{\pi }}\biggl((\frac{\alpha }{i}+\beta )
e^{\frac{2i}{3}[-z(a;k)]^{\frac{3}{2}}+\frac{i\pi}{4}} \nonumber \\
& & -(\frac{\alpha }{i}-\beta )e^{-\frac{2i}{3}[-z(a;k)]^{\frac{3}{2}}
-\frac{i\pi}{4}}\biggr).
\end{eqnarray}
Let us now turn to consider the form of the denominator. This time the
three cases must be treated separately. Let us start with the usual case,
i.e. $k=+1$. The value of the function $z(a;+1)$ when the scale
factor vanishes is given by:
\begin{equation}
\label{z1a0}
z(0;+1)=(\frac{3}{8\pi})^{4/3}(v_1)^{2/3}
(\frac{\rho _{\rm Pl}}{\rho _{\Lambda }})^{2/3}.
\end{equation}
Semi-classical considerations are supposed to be valid only if we
are in a region where $\rho _{\rm Pl}/\rho _{\Lambda }\gg 1$.
 We can therefore
work with the approximation that $z(0;+1)\gg 1$. In that case, we
obtain:
\begin{equation}
\label{num1}
D(k=+1)\approx \frac{\beta }{2\sqrt{\pi }}
[z(0;+1)]^{-1/4}e^{\frac{2}{3}[z(0;+1)]^{3/2}}.
\end{equation}
The case $k=0$ is rather simple since we have $z(0;0)=0$. Therefore, the
denominator can be written as:
\begin{equation}
\label{num0}
D(k=0)=Ai(0)(\alpha +\sqrt{3}\beta ),
\end{equation}
where $Ai(0)=3^{-2/3}/\Gamma (2/3)\approx  0.35502$. Finally, we turn to the
case $k=-1$. Now we have:
\begin{equation}
\label{z-1a=0}
z(0;-1)=-(\frac{3}{8\pi})^{4/3}(v_{-1})^{2/3}
(\frac{\rho _{\rm Pl}}{\rho _{\Lambda }})^{2/3}.
\end{equation}
Therefore, in the semi-classical regime, we have $|z(0;-1)|\gg 1$
and $z(0;-1)<0$. The presence of the minus sign turns out to be
crucial. This time the asymptotic expansion of the Airy functions
for large negative $z$ must be used contrary to the case $k=1$ where the asymptotic
expansion for large positive $z$ has been utilized. Thus, in this limit, the product
$D(k=-1)D^*(k=-1)$ is given by:
\begin{eqnarray}
\label{DD}
& & D(k=-1)D^*(k=-1)\approx \nonumber \\
& & \frac{|\alpha |^2}{2\pi (1+t^2)\sqrt{-z(0;-1)}}
\biggl((1+\rho ^2-2\rho \cos \psi )t^2+2(1-\rho ^2)t
\nonumber \\
& &+1+\rho ^2+2\rho \cos \psi \biggr),
\end{eqnarray}
where $t \equiv \tan \{(2/3)[-z(0;-1)]^{3/2}\}$ and $\beta /\alpha \equiv \rho
e^{i\psi }$. There is now a danger of obtaining a divergence in the current
(see the following section) when the product $D(k=-1)D^*(k=-1)$ vanishes.
The discriminant
of the second order polynomial in $t$ in Eq. (\ref{DD}) is given by:
\begin{equation}
\label{discri}
\Delta=-16\rho ^2\sin ^2\psi.
\end{equation}
Therefore, the polynomial has no real roots
except when $\Delta =0$. This corresponds
to the following cases:
\begin{equation}
\label{deltazero}
\rho =0, \quad \mbox{and/or} \quad \psi =n\pi,
\end{equation}
where $n$ is an integer. Each wavefunction in the minisuperspace is
characterized by the numbers $(\rho ,\psi)$ and thus belongs to a two-dimensional
space. It has been shown by Gibbons and Grishchuk \cite{GG} that this space of the
wavefunctions is in fact a sphere of unit radius. The polar coordinates
$(\theta ,\varphi)$ of a wave function are calculated according to the formulas:
\begin{equation}
\label{coorwf}
\theta =2\tan ^{-1}\biggl(\frac{1}{\rho}\biggr), \quad \varphi =\psi.
\end{equation}
Therefore, the subspace of singular wavefunctions defined by Eqns. (\ref{deltazero})
is just the great circle going through the north and south poles such that
$\varphi =0, \pi $. This shows that almost all the wavefunctions are regular except
those belonging to this subspace. In the case $k=1$, the situation is different since all the
wavefunctions are regular. The Tunneling (Vilenkin's) wave function is such that:
\begin{equation}
\label{State}
\alpha =1, \qquad \beta =i,
\end{equation}
which means that its coordinates are $(\pi/2,\pi/2)$: it is regular. On the other hand,
the Hartle-Hawking state is given by:
\begin{equation}
\label{StateHH}
\alpha =1, \qquad \beta =0,
\end{equation}
i.e. it is represented by the south pole of the sphere and thus belongs
to the subspace of the singular wavefunctions. The divergence occurs when
$t=-1$ (there is only one solution since $\Delta =0$), that is to say
when $\rho _{\Lambda }$ satisfies the following expression:
\begin{equation}
\label{num=0}
\frac{\rho _{\rm Pl}}{\rho _{\Lambda }}=\frac{8\pi ^3}{v_{-1}}
+m\frac{32\pi ^3}{3v_{-1}},
\end{equation}
where $m$ is an integer. This problem is more serious that it might
first appear
because one should not think of $\rho _{\Lambda }$ as being
strictly constant (as we have approximated during the
previous calculations). It rather slowly
varies  in time during the  ``slow roll'' inflationary period. This means
that $\rho _{\Lambda }$ could easily pass through one of these dangerous
values and so cause a divergence in the wave function. Note that this
would be across any chosen  $a=const$ surface, where the semi-classical analysis
should be valid, and so more serious that the previously mentioned factor-ordering
type divergences that can occur  as $a\rightarrow 0$. It is also in
addition to the problems of simply having real wavefunctions, that
do not
allow the current $j$ to be interpreted directly \cite{HP1}.
 This divergent behaviour is like
that found in the context of the quantization of spacetimes which admit a classical
change of signature \cite{signa}.
\par
To know if a divergent value is actually passed through requires a
knowledge of the volume $v_{-1}$.  For $m=0$, we
have $\rho _{\rm Pl}/\rho _{\Lambda }\approx 248/v_{-1}$. Depending on the value
of $v_{-1}$ it could turn out that we are no longer in a regime where the
semi-classical approximation is valid and so the danger could be
ignored as being outside the range of validity. But for
bigger values of $m$, we certainly will be within this region of
 semi-classical validity. For example, if $v_{-1}\approx 0.94$ and $m=3$, one
has $\rho _{\rm Pl}/\rho _{\Lambda }\approx 1316$.
\par
Therefore, it seems that the Tunneling wave function can easily be generalized to describe
the quantum creation of hyperbolic Universes with compact spacelike section whereas
the Hartle Hawking wavefunction leads to important difficulties. It is interesting
to see that considering the quantum creation of compact Universes can lead to
some progress in the debate Tunneling vs no boundary. In the following
 section, we will consider
that the wavefunction is placed in the Tunneling or Vilenkin's
quantum state.
\par
But before ending this section, we would like to discuss the important
case of the Hartle-Hawking wavefunction in more detail. We will concentrate
on the case $k=-1$ versus $k=1$. The first step of the derivation consists
in establishing the Euclidean equations of motion. With the help
of Eqns. (\ref{actgrav}) and (\ref{actmat}), it is easy to see that in the
gauge $\dot{N}=0$ they read:
\begin{eqnarray}
& &\frac{a''}{a}=-\frac{\kappa }{3}\phi '^2-\frac{N^2}{3}\kappa V(\phi ), \\
& &\phi ''+3\frac{a'}{a}\phi '-
N^2\frac{{\rm \partial }V(\phi )}{{\rm \partial }\phi }=0, \\
\label{constraint}
& & \frac{a'^2}{a^2}-\frac{\kappa }{6}\phi '^2-k\frac{N^2}{a^2}+\frac{\kappa N^2}{3}
V(\phi )=0,
\end{eqnarray}
where a prime denotes derivative with respect to $\tau \equiv -it$. The hypersurface
on which we evaluate the wavefunction is the hypersurface such that $\tau =1$
for which we have: $a(\tau =1)\equiv \bar{a}$ and $\phi (\tau =1)\equiv
\bar{\phi }$. The no boundary conditions are [15]:
\begin{equation}
\label{nbconditions}
a(\tau =0)=0, \qquad \frac{{\rm d}\phi }{{\rm d}\tau}(\tau =0)=0.
\end{equation}
In the slow roll regime, the solutions satisfying all the boundary conditions
are found by integrating the Euclidean equations of motion. They are given
by:
\begin{equation}
a(\tau )=\frac{\bar{a}}{\sin \biggl(N\sqrt{\frac{\kappa V}{3}}\biggr)}
\sin \biggl(N\sqrt{\frac{\kappa V}{3}}\tau \biggr), \quad \phi (\tau )=\bar{\phi }.
\end{equation}
The value of $k$ does not appear explicitly in these equations but is
in fact hidden in the algebraic equation satisfied by the lapse function:
\begin{equation}
\sin^2 \biggl(N\sqrt{\frac{\kappa V}{3}}\biggr)=\frac{\kappa \bar{a}^2V}{3k}.
\end{equation}
At this point, one should consider the two cases separately. The case
$k=1$ is standard. The solution of the previous equation can be expressed as:
\begin{equation}
N_{n,k=1}^{\pm}=\sqrt{\frac{3}{\kappa V}}\biggl[(n+\frac{1}{2})\pi
\pm \cos ^{-1}\biggl(\bar{a}\sqrt{\frac{\kappa V}{3}}\biggr)\biggr].
\end{equation}
It is common to consider only the case $n=0$ together with the minus sign, for
a fuller discussion see Ref. [15]. Everything is now present and the Euclidean
action can be calculated along this solution. Then the Hartle-Hawking
wavefunction is given by $\Psi _{\rm HH}(a, \phi)\approx \exp (-S_{\rm E})$ (we
have dropped the bars in order to avoid cumbersome notation) and can be
expressed as:
\begin{equation}
\Psi _{\rm HH}(a, \phi)\approx \exp \biggl\{\frac{6v_1}{\kappa ^2cV}
\biggl[1-\biggl(1-a^2\frac{\kappa V}{3}\biggr)^{3/2}\biggr]\biggr\}.
\end{equation}
In the region where $a\ll a_0$ the previous wavefunction can
be written as $\Psi _{\rm HH}(a, \phi)\approx \exp [(3v_1a^2)/(\kappa c)]$
and we see that this is indeed independent of the scalar field and therefore
that we have picked out the value $\lambda =0$.
\par
Let us now turn to the case $k=-1$. The solution to the algebraic equation
giving the lapse function can be written as:
\begin{equation}
N_{k=-1}=i\sqrt{\frac{3}{\kappa V}}
\sinh ^{-1}\biggl(\bar{a}\sqrt{\frac{\kappa V}{3}}\biggr) .
\end{equation}
Interestingly enough the solution is now unique. The explicit expression of the
scale factor can be easily deduced. It reads:
\begin{equation}
a(\tau )=\sqrt{\frac{3}{\kappa V}}\sinh \biggl[\sinh ^{-1}\biggl(
\bar{a}\sqrt{\frac{\kappa V}{3}}\biggr)\tau \biggr].
\end{equation}
In the same manner, we can evaluate the Euclidean action and therefore
find the corresponding no boundary state. One obtains:
\begin{equation}
\Psi _{\rm HH}(a, \phi)\approx \exp \biggl\{i\frac{6v_{-1}}{\kappa ^2cV}
\biggl[1-\biggl(1+a^2\frac{\kappa V}{3}\biggr)^{3/2}\biggr]\biggr\},
\end{equation}
where we have again suppressed the bars. In the limit where the scale
factor goes to zero, this wavefunction can be written
as $\Psi _{\rm HH}(a, \phi)\approx \exp [-(3iv_{-1}a^2)/(\kappa c)]$. It
confirms that we have picked out a vanishing separation constant. This also
shows that in the Euclidean region, the wavefunction is oscillatory. One
could easily find the no boundary wavefunction in the classical region
with the WKB matching method. Of course, we would recover the fact that
the denominator is now a trigonometric function which was the main reason
for arguing that the Vilenkin state is preferable. In fact we can
go a step further and say that the very concept of a no boundary wavefunction
is in danger by this kind of analysis. Indeed, from Eq. (\ref{constraint}) we
see that the no boundary conditions (\ref{nbconditions}) imply that:
\begin{equation}
\frac{1}{N^2}\biggl(\frac{{\rm d}a}{{\rm d}\tau}\biggr)^2=k.
\end{equation}
In order for the metric ${\rm d}s^2=N^2{\rm d}\tau ^2+a^2(\tau ){\rm d}\Omega _3^2$
to be Euclidean and regular at $\tau =0$ we need $a(\tau )\approx N\tau $
which implies $k=+1$! In the case $k=-1$, it is easy to see that we have
in fact $a(\tau )\approx -iN\tau $. This means that ${\rm d}s^2\approx N^2(
{\rm d}\tau ^2-\tau ^2{\rm d}\Omega _3^2)$. The corresponding manifold is
no longer Euclidean. In addition, the metric found above
is very reminiscent of that used by Hawking and Turok [9] in their
analysis. Therefore it seems that there exists an interesting link between the
approach advocated here and the one of Ref. [9]. This requires further
studies which are beyond the scope of the present article.
\par
To end this section, let us come back to the question of the regularity
of the wave function. We have just shown that the requirement
$|\Psi (a;k)|< \infty$ everywhere in the minisuperspace and for any $k$ favours
the Tunneling wavefunction over the Hartle Hawking state. On the other hand
 requiring that the wavefunction be regular as $a\rightarrow 0$ for every factor
ordering leads to a Hartle Hawking state for the $k=1$ case \cite{australia}.
 Therefore, the two requirements
are strictly not compatible if the analysis was extended to $p>1$.
 However, we have previously mentioned that
in more realistic models
the actual presence of non-inflationary matter could also complicate
the adoption of the HH wavefunction.
\par
Regardless of these complications, and because factor
ordering problems, which occur
 in the unrealistic limit $a\rightarrow 0$, appear
  less serious than the divergences found
in this section for $k=-1$ models, we will next consider the
predictions with these Tunneling wavefunctions.

\section{Predictions}

In this section, we address the problem of computing physical
predictions from the previously obtained wavefunctions. It is well
known that, due to the fact of obtaining non-normalizable wavefunctions, this
is a difficult task. In Ref. \cite{Halli}, a method to overcome this
problem has been proposed. The idea is to use the current
defined by Eq. (\ref{defcurrent}) in the WKB regime. This leads to
well-defined probabilities. Using the asymptotic form of the wavefunction
given by Eq. (\ref{wkbnum}), the following expression for the current, valid
for any $k$, can be found:
\begin{equation}
\label{currentanyk}
j(k)=\frac{2}{\pi }(\frac{3}{8\pi})^{1/3}\frac{v_k^{2/3}}{D(k)D^*(k)}
(\frac{\rho _{\rm Pl}}{\rho _{\Lambda }})^{-1/3}.
\end{equation}
As expected this expression no longer depends on the scale
factor. Going further requires a knowledge of $D(k)$ so
we must now treat the three cases separately. Let us first consider
the case $k=1$. The expression of $D(k=1)$ can be deduced from
Eq. (\ref{num1}). This leads to:
\begin{equation}
\label{currentwkb1}
j(k=+1)=\frac{3v_1}{\pi }e^{-\frac{3v_1}{16 \pi ^2}
\frac{\rho _{\rm Pl}}{\rho _{\Lambda }}}.
\end{equation}
We recover the well-known expression for the closed case. The coefficient
of proportionality is not of great interest since
the probability distribution is not normalizable and instead must be used
to calculate conditional probabilities, see below. The ratio
$\rho _{\rm Pl}/\rho _{\Lambda }$ only appears in the argument
of the exponential function. This is because the factor $[z(0;+1)]^{-1/2}$
in the $D(k=1)D^*(k=1)$ term cancels exactly the
term $(\rho _{\rm Pl}/\rho _{\Lambda })^{-1/3}$ in Eq. (\ref{currentanyk}).
The expression of the current in the case $k=0$ can be established
from Eq. (\ref{num0}). It reads:
\begin{equation}
\label{currentwkb0}
j(k=0)=\frac{1}{4\pi Ai^2(0)}(\frac{3}{\pi })^{1/3}v_0^{2/3}
(\frac{\rho _{\Lambda }}{\rho _{\rm Pl}})^{1/3}.
\end{equation}
Since $D(k=0)$ is just a constant, this does change the dependence in
$(\rho _{\rm Pl}/\rho _{\Lambda })^{-1/3}$ of Eq. (\ref{currentwkb1}).
 Finally,
we turn to the case $k=-1$. From Eq. (\ref{DD}), we have
that $D(k=-1)D^*(k=-1)=(1/\pi )[-z(0;-1)]^{-1/2}$. This time no exponential
function appears as  was the case for $k=1$. This is due to the fact
that we have chosen the boundary condition  that $D(k=-1)$ simply
be a phase. However, now for the case $k=-1$, the
 factor $\rho _{\rm Pl}/\rho _{\Lambda }$ simply cancels
out. As a consequence, one finds:
\begin{equation}
\label{currentwkbm1}
j(k=-1)=\frac{3v_{-1}}{4\pi }.
\end{equation}
The current turns out to be independent of $\rho _{\Lambda}$.
\par
As already mentioned, the current can now be used to
calculate conditional probabilities. For example, the probability of
having an initial value $\phi_{\rm i}$ of the scalar field, greater than the
value  needed to solve the problems of standard
cosmology, $\phi _{\rm suf}$, knowing that $0<\phi <\phi _{\rm sup}$ can be
allowed. In the last inequality $\phi _{\rm sup}$ is the value at which
semi-classical considerations cease to be valid, i.e. when the
potential reaches the Planck scale, $\rho _{\rm Pl}=\rho _{\Lambda }
=V(\phi )=m_{\rm Pl}^4$ in the Planck system of units. Also, the fact
that the minimal value for $\phi $ is zero is not a problem here because
the wave function is the Tunneling one. This would no longer be true
if the state were the HH one. We will evaluate the conditional
probabilities for the prototype chaotic inflationary scenario,
with scalar potential of the form
 $V(\phi )=(\lambda /4!)\phi ^4$. In this context, the initial value
of the field necessary
to get $N$ e-folds is $\phi _{\rm i}=\sqrt{(N+1)/\pi}m_{\rm Pl}$. If
we consider that sufficient inflation is obtained when $N=60$ then we
find that the scalar field has to start at
$\phi _{\rm i}\equiv \phi _{\rm suf }=4.4 m_{\rm Pl}$. The value of
$\phi _{\rm sup}$ is given
 by $\phi _{\rm sup}=24^{1/4}\lambda ^{-1/4}m_{\rm Pl}$. For
chaotic inflation,
one has $\lambda \sim 10^{-15}$ in order
 to reproduce the value of the quadrupole
of the Cosmic Microwave
Background Radiation anisotropy, $Q_{\rm rms-PS}$, measured by the
COBE satellite, i.e. $Q_{\rm rms-PS}\approx 18 \times 10^{-6}$ K \cite{cobe}. This
implies $\phi _{\rm sup}\approx 1.2\times 10^4 m_{\rm Pl}$. The probability to have
sufficient inflation in these three cases is computed according to the
formula \cite{Halli}:
\begin{equation}
\label{Proba1}
P(k;\phi _{\rm i}>\phi _{\rm suf}|0<\phi _{\rm i}<\phi _{\rm sup})\equiv P(k)=
\frac{\int _{\phi _{\rm suf}}^{\phi _{\rm sup}}j(k){\rm d}\phi }
{\int _0^{\phi _{\rm sup}}j(k){\rm d}\phi }.
\end{equation}
It will be more convenient in the
 following to rewrite the previous definition
of the conditional probabilities as $P(k)=1-R(k;N,\lambda)$ where $R(k;N,\lambda )$
is given by:
\begin{equation}
\label{defR}
R(k;N,\lambda)\equiv \frac{\int _0^{\phi _{\rm suf}}j(k){\rm d}\phi }{
\int _{0}^{\phi _{\rm sup}}j(k){\rm d}\phi }
\end{equation}
We are now going to calculate $R$ in the three cases. Let us start with the case
$k=1$. The current $j(1)$ is given by Eq. (\ref{currentwkb1}). Using the change of
variable $u\equiv 9m_{\rm Pl}^4/(\lambda \phi ^4)$ and the formula (3.381.6) of
Ref. \cite{GR} one can easily show that the denominator
is given by $(24/9)^{5/8}e^{-3/16}W_{-\frac{5}{8},-\frac{1}{8}}(9/24)$ where
$W_{\mu ,\nu}$ is a Whittaker function. Since $W_{-\frac{5}{8},-\frac{1}{8}}(9/24)
\approx 0.52$ we find that the denominator
is approximately equal to $0.80$. In the same manner, the numerator is
equal to:
\begin{equation}
\label{denocurrent}
\biggl[\frac{9\pi ^2}{\lambda (N+1)^2}\biggr]^{-5/8}e^{-\frac{9\pi ^2}{2\lambda (N+1)^2}}
W_{-\frac{5}{8},-\frac{1}{8}}\biggl(\frac{9\pi ^2}{\lambda (N+1)^2}\biggr).
\end{equation}
Since the parameter $\lambda $ appears at the denominator of the argument
of the Whittaker function, this has a very large value. Therefore, we can use the
asymptotic expansion of the Whittaker function for large values of its
argument given by \cite{GR}:
\begin{equation}
\label{Whitexp}
\lim _{|z|\rightarrow \infty}W_{\mu ,\nu}(z)=e^{-\frac{z}{2}}z^{\mu }\biggl(1+
{\cal O}(\frac{1}{z})\biggr).
\end{equation}
Thus, the value of the function $R(1;N,\lambda)$ can be expressed as:
\begin{equation}
\label{R1}
R(1;N,\lambda )\approx 0.0045\lambda ^{\frac{5}{4}}(N+1)^{\frac{5}{2}}
e^{-\frac{9\pi ^2}{\lambda (N+1)^2}}.
\end{equation}
This expression is valid for any value of $N$ and $\lambda $ provided that
$\lambda $ is a small number. Putting $N=60$ and $\lambda =10^{-15}$, we find that
$R(1;60,10^{-15})\approx 10^{-10^{13}}$, an extremely small number which has its
origin in the presence of the
parameter $\lambda $ in the exponential factor. We conclude that
$P(1)\approx 1$. The calculation of $P(0)$ is easier. Using Eq. (\ref{currentwkb0}),
we find that:
\begin{equation}
\label{R0}
R(0;N,\lambda )=24^{-\frac{7}{24}}\biggl(\frac{N+1}{\pi }\biggr)^{\frac{7}{6}}
\lambda ^{\frac{7}{12}}.
\end{equation}
This gives $R(0;60,10^{-15})\approx 1.5\times 10^{-9}$ leading to $P(0)\approx 1$
again. Finally, the case $k=-1$ is straightforward. Using Eq. (\ref{currentwkbm1}),
we can easily establish that:
\begin{equation}
\label{Rm1}
R(-1;N,\lambda )=24^{-\frac{1}{4}}\biggl(\frac{N+1}{\pi }\biggr)^{\frac{1}{2}}
\lambda ^{\frac{1}{4}}.
\end{equation}
This results in $R(-1;60,10^{-15})\approx 1.67\times 10^{-4}$. Therefore, one can
say that $P(-1)\approx 1$. The conclusion is that, in the three cases, the probability
turns out to be close to one. This means that sufficient inflation is a
prediction of the Tunneling wavefunction whatever the
value of $k$ is. In this respect, the three cases are equally compatible with
there being a near definite prediction of inflation occurring. This is actually an
significant improvement over classical notions of whether inflation will
occur. Which, due to an infinite divergence over an arbitrary scale factor, gives
an ambiguous prediction even for apparently inflationary potentials
\cite{Coule}. We note however than the probability is closer to one in the
case $k=1$ than in the cases $k=0,-1$.
\par
We can pursue this reasoning a step further by also allowing $k$
to be quantum (q) variable. Such a possibility could be a first step
in modeling quantum fluctuations in the geometry. Unlike the
classical case where the curvature is simply a fixed (c) number
constant we are now assuming that the curvature also is in a quantum
ensemble.  The conditional
probability for having a given $k$ knowing that $k=0,\pm 1$ can be
calculated for a fixed value of the scale factor. This probability is
formally defined
according to the equation:
\begin{equation}
\label{Proba2}
P(a;k)\equiv \frac{|\Psi (a;k)|^2}{\sum _{l=0,\pm 1}|\Psi (a;l)|^2}.
\end{equation}
The use of this definition requires some comments. Let us first recall
how the definition of Eq. (\ref{Proba1}) can be justified. Let
${\cal M}$ be the minisuperspace and
 ${\cal M}_{\rm WKB}\in {\cal M}$ the region
of the minisuperspace where the wavefunction can be well approximated by the
WKB wavefunction. In non relativistic quantum mechanics, the time
component of the current, $|\Psi |^2$,  gives the probability density
function. Since the wavefunction is normalizable, the probability of finding
a particle in the interval $[a,b]$ is given
by $\int _a^b{\rm d}x |\Psi |^2/\int _{-\infty }^{\infty }{\rm d}x |\Psi |^2$. In
quantum cosmology, the current
 is positive definite only in the WKB regime. In
this regime, $\Psi (a;k)$ is not normalizable,
i.e. $\int _{{\cal M}_{\rm WKB}}{\rm d}q^{\alpha }|\Psi |^2$, where
${\rm d}q^{\alpha }$ is the volume in ${\cal M}$, is not finite.
 This does not
mean that $\int _{\cal M}\mu (q^{\alpha }){\rm d}q^{\alpha }|\Psi |^2$ is infinite
since the measure $\mu (q^{\alpha })$ is {\em a priori} not known.
 In the WKB regime, the
wavefunction is by definition peaked over ${\cal M}_{\rm WKB}$. Therefore, one
has $\int _{\cal M}\mu (q^{\alpha }){\rm d}q^{\alpha }|\Psi |^2
\approx \int _{{\cal M}_{\rm WKB}}{\rm d}q^{\alpha }|\Psi |^2$. Thus, the same
rule as in ordinary quantum mechanics says that the probability of
finding the system in the region ${\cal R}\in {\cal M}_{\rm WKB}$ is given
by $\int _{\cal R}{\rm d}q^{\alpha }|\Psi |^2/
\int _{{\cal M}_{\rm WKB}}{\rm d}q^{\alpha }|\Psi |^2$. Eq. (\ref{Proba1}) is a
special case of this more general formula.
\par
The interpretation of Eq. (\ref{Proba2}) is roughly  the same.
Although it is, of course, more contentious
to apply this reasoning to obtain
the geometry, here represented by $k$, compared to how we previously
obtained  the initial matter component $\phi$. The matter calculation is
a fairly straightforward adaptation, as stated above,
 of usual quantum mechanics reasoning whereas
the quantization of the geometry might require more extensive
alterations to quantum mechanics.
 We will proceed with the notion that $k$
is not initially fixed but is rather undefined in a quantum state
with ``equipartition'' among all possible $k$.  In the context of the
histories approach of quantum mechanics, $P(a;k)$ represents the probability
that the Universe ``choose'' one of these possible histories.
 It will not come
as a surprise that the Universe goes down the lowest potential case.
\par
We would like
to emphasize how $P(a;k)$ differs from the notion of the
 probability of a change of topology once
the Universe has been created, i.e.
the probability of having, for example, $k=1$ for some value of the
scale factor and then $k=0$ for another (bigger) value. If
topology changes are required from one classical model to another
then an explicit time dependent curvature should be introduced.
Such a construction with  ${\rm d}k/{\rm d}t\neq 0$
has been explicitly made in Ref. \cite{Netal} and it has been shown that, in this
case, a passage to a midi-superspace description is mandatory.
However in the present example we are assuming that the curvature
is first a quantum variable and so not fixed
 in a particular classical state. What classical curvature state the
 universe first evolves to is our present concern, not whether topology
 changes still occur once this classical state is first achieved.
 As
the universe becomes increasingly classical, by for example gaining
energy from falling down the potential, then the curvature will
no longer be in a quantum superposition but will become
increasingly in a specific classical state. How this ``measurement''
 takes place
will be a rather complex process and likely to depend on
 quantum mechanical interpretational questions. But for our present
 purposes,  this rough notion that the curvature will eventually
 ``crystallize out'' into a classical state should
suffice.
\par
On the following figures, the evolution of the three probabilities
as a
function of the scale factor are displayed. The first figure is for
the case  $\rho _{\rm Pl}/\rho _{\Lambda }=10$. This means
that the dimensional quantity $a_0$ is equal
to $a_0\approx l_{\rm Pl}$. Let us
 also recall that we have taken $v_1=2\pi ^2$,
$v_0=1$ and $v_{-1}\approx 0.94$. The choice of the volume can influence
the behaviour of the probabilities for small scale factors but in the limit
of big scale factors, they are mainly determined by the
ratio $\rho _{\rm Pl}/\rho _{\Lambda }$.
\begin{figure}
\begin{center}
\leavevmode
\hbox{%
\epsfxsize=8cm
\epsffile{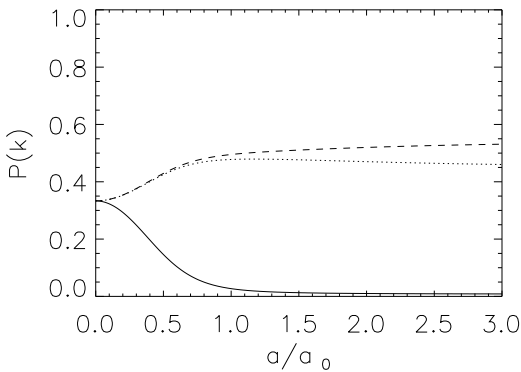}}
\end{center}
\caption{Probabilities $P(k)$ for $\rho _{\rm Pl}/\rho _{\Lambda }=10$.
The solid line represents the case $k=1$, the dotted line is the $k=0$ case
and dashed line is the $k=-1$ case.}
\label{proba10}
\end{figure}
The second figure represents the case where $\rho _{\rm Pl}/\rho _{\Lambda }=100$. This
corresponds to $a_0\approx 3.5 l_{\rm Pl}$.
\begin{figure}
\begin{center}
\leavevmode
\hbox{%
\epsfxsize=8cm
\epsffile{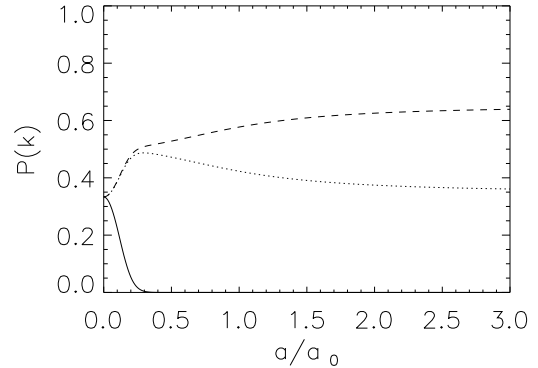}}
\end{center}
\caption{Probabilities $P(k)$ for $\rho _{\rm Pl}/\rho _{\Lambda }=100$.
The solid line represents the case $k=1$, the dotted line is the $k=0$ case
and dashed line is the $k=-1$ case.}
\label{proba100}
\end{figure}
Finally, the third and last figure represents the case
$\rho _{\rm Pl}/\rho _{\Lambda }=1000$ which corresponds
to $a_0\approx 10.9l_{\rm Pl}$ as already mentioned. This
case could be considered as the most realistic one since inflation takes
place at an energy comparable to the GUT scale, i.e. $\rho _{\Lambda }\approx 10^{16}
\mbox{GeV}$.
\begin{figure}
\begin{center}
\leavevmode
\hbox{%
\epsfxsize=8cm
\epsffile{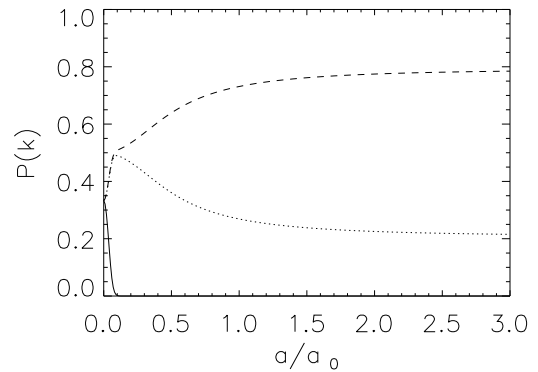}}
\end{center}
\caption{Probabilities $P(k)$ for $\rho _{\rm Pl}/\rho _{\Lambda }=1000$.
The solid line represents the case $k=1$, the dotted line is the $k=0$ case
and dashed line is the $k=-1$ case.}
\label{proba1000}
\end{figure}
Let us now comment on these figures in more detail. In each case, we can point out the
following features. At vanishing scale factor, the probabilities are all assumed
equal to $1/3$ for any value of $k$. Provided the probabilities are roughly
equal the actual values will not affect the predictions
significantly. In the region
where the scale factor is still of order $a_0$, the behaviour
of the probabilities are rapidly evolving with $a$. Finally,
 when the scale factor
becomes large in comparison with $a_0$, the probabilities tend to a constant
value which depends on the ratio $\rho _{\rm Pl}/\rho _{\Lambda }$.
 This suggests some
interesting ideas. For example,
 the fact that the probabilities remain similar in the
region $a<a_0$ could mean that here the topology can easily
fluctuate due to quantum effects. On the other hand, when the scale factor
is such that $a\gg a_0$, the probabilities differ significantly and
are almost constant as a function of $a$. This can be
easily understood if one looks at the asymptotic behaviour of
$|\Psi (a;k)|^2$ when $a$ becomes large. Using Eq. (\ref{solpsi}), one finds
that:
\begin{eqnarray}
\label{k=1}
& &|\Psi (a;k=+1)|^2 \approx
\frac{4}{a}(\frac{3}{8\pi })^{1/2}(\frac{\rho _{\rm Pl}}{\rho _{\Lambda }})^{1/2}
e^{-\frac{3v_1}{16\pi ^2}
\frac{\rho _{\rm Pl}}{\rho _{\Lambda }}}, \\
\label{k=0}
& &|\Psi (a;k=0)|^2 \approx
\frac{1}{4\pi Ai^2(0)}(\frac{8\pi }{3})^{1/6}
\frac{1}{av_0^{1/3}}(\frac{\rho _{\rm Pl}}{\rho _{\Lambda }})^{1/6}, \\
\label{k=-1}
& &|\Psi (a;k=-1)|^2 \approx \frac{1}{a}(\frac{3}{8\pi })^{1/2}(\frac{\rho _{\rm Pl}}{
\rho _{\Lambda }})^{1/2}.
\end{eqnarray}
In each case $|\Psi (a;k)|^2$ behaves as $1/a$ and therefore $P(a;k)$ becomes
independent of the scale factor when $a\gg a_0$. Fluctuations in topology
are more likely, regardless of the scale factor, provided the energy density
is near Planck values. But as the value of $\rho_{\Lambda}$ reduces this
rapidly becomes less likely. When
$\rho _{\rm Pl}/\rho _{\Lambda }\gg 1$, it is clear from the figures
that a definite prediction can be made since one of the probabilities becomes
equal to one:
\begin{equation}
\label{Probresult}
P(-1)\approx 1, \quad P(0)\approx 0, \quad P(+1)\approx 0.
\end{equation}
The fact that the greater the ratio $\rho _{\rm Pl}/\rho _{\Lambda }$
is, the
better the prediction is [i.e. the closer to one $P(-1)$ is] means that when
inflation takes place at energy well below the Planck scale (typically the GUT scale
seems to be the most physical case), topology changes
become strictly forbidden as the scale factor becomes large. In
practice as the scalar field rolls down the potential the
effective cosmological constant $\rho_{\Lambda}$ reduces and topology changes
become increasingly unlikely.
\par
Finally, let us comment on the prediction itself. It says, since
an inflationary phase is expected to give a final $a\stackrel{>}{\sim}
10^{30}l_{\rm Pl}$, that
  our Universe is most likely to  be open, or at worse flat, a
quite interesting statement indeed. Bearing in mind that
 we have not tried to allow for the
more numerous hyperbolic topologies corresponding to $k=-1$, we have
probably still underestimated  this overwhelming probability
 of obtaining an open universe.   However, it does not say
 that $\Omega _0$ is necessarily much less than unity and so could
actually be distinguished from
$\Omega _0=1$ by observations. This is because we have
 seen that even in the
open case, sufficient inflation is a prediction of the Tunneling wavefunction.
Thus, inflation will probably drive $\Omega _{\rm ini}<1$ to a value
very close to one even at the present age of the universe.
 The actual form of the scalar potential is crucial for determining
such properties, a quantum description
 alone  does not  mandate
such values for $\Omega _0$. Once the inflationary phase finishes the
matter behaves effectively like that of  dust \cite{Coule} so now
obeying the strong-energy condition. Eventually the curvature term
will start again to dominate the dynamics, with the open model
expanding infinitely into the future. Curvature can play an
important
role at both the beginning and end of the universe's evolution
because of the drastically different change in the behaviour of
matter:
from being a cosmological constant to becoming like dust. Any
small residual cosmological constant would also come to dominate at later
times.
Recent development, although still contentious,  have suggested
 such a $\Lambda$ term  is necessary to explain
the Supernova data \cite{SNIa}, but such a value
is still extremely small $\rho_{\Lambda}
\sim 10^{-120}$. Understanding such fine detail while quantum
cosmology takes a rather ``broad brush'' approach to calculating the
various quantities remains a difficulty. We would just remind
readers that the wormhole, and related  mechanisms, that were suggested should
predict $\Lambda=0$ exactly \cite{Col}, could instead give other values which
are only approximately zero -see e.g.\cite{Cline}.

\section{Conclusions}

In general relativity the global topology has to be imposed as an
initial condition. In quantum cosmology most studies have
concentrated on elliptic space which  for the FLRW model is
simply  a closed
universe. However, hyperbolic and flat spaces are mathematically
speaking more numerous and many are also compact. These give rise
to geometries that are locally described by the FLRW metric with
$k=-1,0$.
For the simple DeSitter model the $k=-1,0$ cases are
no longer distinguished by the presence of a forbidden or Euclidean region
at small scale factors.

Because the Euclidean nature of the model is now absent  it might seem
that the smooth geometric picture of the Hartle-Hawking
no-boundary proposal is a serious loss.
But a major weakness of ``quantum creation of the universe''
 ideas   is that the
forbidden region is anyway eroded by the presence of
strong-energy satisfying matter. Such matter will generally
prevent the Euclidean
nature of the model for small scale factors.
 But on general grounds such matter should be
present,if only because of ``zero-point'' fluctuations. A
Lorentzian region will tend to form anyway as $a\rightarrow 0$.

With a scalar field source a possible problem still remains since
a singularity caused by the kinetic energy of the scalar
field blowing up could occur. The kinetic energy of the scalar field behaves as
a `stiff' equation of state. This causes an effective $\sim -a^{-2}$
term in the WDW potential that would tend to create an infinite
`wiggliness' in the wavefunction as $a\rightarrow 0$. By ensuring
that the wavefunction be independent of the matter as
$a\rightarrow 0$ this possible problem is avoided. But one must
bear in mind that one is removing the singularity by fiat, one
should not claim that the process of quantization alone is
achieving this, as many studies erroneously assume. Likewise in
the usual closed models the Hartle Hawking boundary condition
only gives a Euclidean region by imposing that the matter
fields are not allowed to dominate as $a\rightarrow 0$.

 Once it is accepted that a Lorentzian region is
anyway present for the smallest  scale factors it is no longer  a major
fundamental difference whether one then has a forbidden region
away from the origin.
In the flat and open cases the forbidden region indeed goes away
and we have argued that these cases are in a sense more favourable.
  Further, in the open case the requirement that the wavefunction
  should remain finite at arbitrary $a$ allows boundary conditions that
  include  Vilenkin's ``outgoing
only'' but not the no boundary choice.

  For the specific Tunneling boundary condition, comparisons between models that only differ by their
 value of $k$ can be made. If the initial curvature is given by a quantum
 ensemble, so allowing any possible $k$,
  then one can conclude that open universes are
 more likely. Although the models considered have only a
 cosmological constant as their matter source. This seem to
 contradict the notion that for DeSitter space ``all curvatures
 are equivalent''. But because we consider the evolution from
 $a=0_{+}$ and regularized the wavefunction around this point we
 have derived ``propagators'' to go to arbitrarily large scale
 factors. The universe is given a choice from its conception which path
 to follow. In the closed model one can think of the universe being
  ``held up'' waiting to  tunnel through the barrier. While the open
  models gain a ``push'' in falling down a steeper WDW potential
  compared to the flat case. Once the scale factor becomes
  sufficiently large the possibility of topology change depends
  entirely on the energy density of the scalar field driving
  inflation. As this reduces below Planck values the prediction
  rapidly gives that an open $k=-1$ universe is favoured.
  Interestingly at small scale factors and/or Planck energy
  densities the universe might ``pin ball'' between various
  possibilities before settling into one final curvature. The
  ``no hair'' property of DeSitter space, of having finite causal horizon,
   might also allow various
  regions to develop different curvatures.  Although the open case
  would still dominate pockets of closed curvature could also
  be created. This could depend on quantum
  interpretations, whether ``many worlds'' or
  ``single-history'' quantum theories are possible cf. \cite{Page}.

Regardless of the actual value of curvature inflation is predicted
in all cases. This resolves an ambiguity in classical measures
of inflation. If quantum cosmology did nothing else but gave a
definite predictions for inflation to occur it would be very significant.
The only drawback is that the scalar potential must be chosen to be of
the correct shape, just as the potential of the Hydrogen atom
has to be provided before one does any quantum mechanics. The
potential is further constrained by the need to create sufficiently
small fluctuations and gravitational waves: roughly speaking
inflation should occur at GUT $\sim 10^{14}\mbox{GeV}$ energy scales
\cite{RSV,AW}. Ultimately inflation is a classical phenomena that can not be driven
by quantum notions alone.  We have modeled curvature as being described
by a quantum variable and subject to notion  of probability analogous to
the usual initial matter distribution calculation.
There is the worry that in
trying to determine the curvature, which is also a part of the
potential, we are going beyond the scope of what quantum mechanics
should be used to  predict. We have assumed that the universe is free to
take the path of least resistance i.e. follow  the open case,
 but maybe there
is actually no freedom in this choice and it is pre-ordained what
curvature the universe should take
before the universe comes into existence. This is a rather deep
problem that affects quantum cosmology in general, what variables
are free to roam  and which are fixed externally imposed constants ?
Is there a classical scaffolding surrounding the initial universe
or is every variable initially a quantum variable?
Quantum cosmology seems ambiguous why certain variables (e.g. $\phi$)
 are given
by distribution functions, although constrained by the
chosen boundary conditions, while others (e.g. $k$)
 are imposed with no longer apparent quantum uncertainty. Recall
that all classical notions seem suspect as the Planck epoch is
approached cf. \cite{Sch}. But in the
meantime it seems that, in admittedly simplistic models, different
curvatures can be considered and an argument made that the open
case is favoured. If the curvature is initially fixed and so not
subject to quantum uncertainty, then one can still argue that open
universes are as possible as closed ones with Tunneling like
boundary conditions. Understanding better the measure of possible
topologies would seem the next helpful step to see if arguments can be made to
favour a specific choice cf.\cite{Hartle}

In summary, we have considered quantum cosmological models with
arbitrary curvature. Although the models are all compact they are
all possible candidates for quantum creation of the universe
models. Unlike always starting from closed models which have a
forbidden region at small scale factors, one can work directly with
the curvature of one's choice. Compared to recent instanton
methods of creating an open universe from an initial closed one, one cuts
out the unnecessary closed stage. One has the further prediction that open
universes are favoured followed by flat ones provided in some sense
the universe has the choice of deciding its curvature. The closed universe
case is strongly suppressed in comparison. One can still obtain $\Omega _0
\approx 1$ in all cases since a long period of inflation is
strongly predicted. Unfortunately this large inflationary period would
appear to
wipe out any interesting ``multiple images'' due to topological effects that
are presently being searched for. Only for lesser amount of inflation would
such effects be apparent in the patterns of Cosmic
 Background radiation or multiple galaxy images -see \cite {Topo}.
 But interestingly  with an inflationary
  matter source, the curvature still dominates
the dynamics  both at the
beginning and the end of the universe, this seems reasonable that
the end of the universe should reflect its origins. This is unlike
the conventional big bang model where the curvature only dominates
in the far future and ``only matter matters'' during its initial phase.

\acknowledgements
It is a pleasure to thank N. Pinto Neto for useful exchanges and comments.
We are also grateful to D. Wiltshire for various  helpful
remarks  concerning the factor ordering issue.

\end{document}